# The open access coverage of OpenAlex, Scopus and Web of Science

Marc-André Simard[*], Isabel Basson[**], Madelaine Hare[***], Vincent Larivière[****] and Philippe Mongeon[*****]

[*]*marc-andre.simard.1@umontreal.ca*
École de bibliothéconomie et des sciences de l'information, Université de Montréal, Pavillon Lionel-Groulx, 3150 rue Jean-Brillant, Montréal, Québec, H3T 1N8 (Canada)

[**]*isabel.basson@umontreal.ca*
École de bibliothéconomie et des sciences de l'information, Université de Montréal, Pavillon Lionel-Groulx, 3150 rue Jean-Brillant, Montréal, Québec, H3T 1N8 (Canada)

[***]*maddie.hare@uottawa.ca*
School of Information Studies, University of Ottawa, Desmarais Building, 11th Floor, 55 Laurier Ave. East, Ottawa, Ontario, K1N 6N5 (Canada)

[****]*vincent.lariviere@umontreal.ca*
École de bibliothéconomie et des sciences de l'information, Université de Montréal, Pavillon Lionel-Groulx, 3150 rue Jean-Brillant, Montréal, Québec, H3T 1N8 (Canada)

Observatoire des sciences et des technologies, Centre interuniversitaire de recherche sur la science et la technologie (CIRST), Université du Québec à Montréal, Pavillon Paul-Gérin-Lajoie, 1205 rue Saint-Denis, Montréal, Québec, H2X 3R9 (Canada)

[*****]*pmongeon@dal.ca*
Department of Information Science, Dalhousie University, Kenneth C. Rowe Building, 6100 University Avenue, Halifax, Nova Scotia, B3H 4R2

Centre interuniversitaire de recherche sur la science et la technologie (CIRST), Université du Québec à Montréal, Pavillon Paul-Gérin-Lajoie, 1205 rue Saint-Denis, Montréal, Québec, H2X 3R9 (Canada)

## Abstract

Diamond open access (OA) journals offer a publishing model that is free for both authors and readers, but their lack of indexing in major bibliographic databases presents challenges in assessing the uptake of these journals. Furthermore, OA characteristics such as publication language and country of publication have often been used to support the argument that OA journals are more diverse and aim to serve a local community, but there is a current lack of empirical evidence related to the geographical and linguistic characteristics of OA journals. Using OpenAlex and the Directory of Open Access Journals as a benchmark, this paper investigates the coverage of diamond and gold through authorship and journal coverage in the Web of Science and Scopus by field, country, and language. Results show their lower coverage in WoS and Scopus, and the local scope of diamond OA. The share of English-only journals is considerably higher among gold journals. High-income countries have the highest share of authorship in every domain and type of journal, except for diamond journals in the social sciences and humanities. Understanding the current landscape of diamond OA indexing can aid the scholarly communications network with advancing policy and practices towards more inclusive OA models.

## Introduction

The Budapest Open Access Initiative (2002) advocated for a new generation of open access (OA) journals that would rely on alternative sources of funding to ensure free and unrestricted access to scientific literature. Article processing charges (APCs) for authors, initially developed in the early 2000s by new OA publishers such as the Public Library of Science (PLOS), emerged as a commonly used source of revenue for OA publishing (PLOS, 2022), particularly among for-profit publishers (Butler et al., 2023; Siler & Frenken, 2020). APCs are sometimes considered a necessary evil of OA publishing, with one of the main justifying narratives being that "someone has to pay" for the costs of publication (Meadows, 2014). Thus, the APC-based models of OA publishing transfer the burden of cost to the researchers who produce the work, their institutions, funding bodies, and governments. APCs have been criticized for contributing to the exclusion of and heightening inequalities among early career researchers, researchers from low-income countries or specific disciplines, and other groups that lack representation in the scientific research system (Burchardt, 2014; Cabrerizo, 2022; Klebel & Ross-Hellauer, 2022; Kwon, 2022; Momeni et al., 2023; Olejniczak & Wilson, 2020; Ross-Hellauer, 2022; Ross-Hellauer et al., 2022; Smith et al., 2021). APCs have also garnered a lot of criticism among the scientific community for their high prices, particularly among major Western for-profit publishers where APCs are typically in the $3,000 USD - $5,000 USD range and can go as high as $11,500.00 USD (Butler et al., 2023; Else, 2020; Simard, 2021).

Despite the wide adoption of APC-based publication models, other models that are free for both readers and authors have existed for decades across the globe (e.g., Érudit, OpenEdition, SciELO). These types of models were recently rebranded by OA advocates as "diamond OA" (or sometimes "platinum OA) to promote "non-commercial publishing models for Open Access" (cOAlition S, 2020). While diamond journals often rely on various sources of funding, including in-kind support, voluntary labour, university and government grants, crowdfunding, donations, memberships, and shared infrastructure, many of them still struggle with breaking even (Bosman et al., 2021).

Despite concerns about their quality, sustainability, and scalability (Alperin, 2022; Bosman et al., 2021; Suber, 2009), diamond journals could represent an alternative to expensive gold journals in a context where fees asked by prestigious OA journals from for-profit publishers have been steadily increasing over time (Butler et al., 2023). However, diamond OA journals are not well represented in widely used bibliometrics databases such as Web of Science (WoS) and Scopus (Bosman et al., 2021; Khanna et al., 2022), potentially hindering their findability of readers and attractiveness for authors, bibliometric analyses, and the ability of funding and policy bodies to set up and sustain OA policies, infrastructure, and services (Becerril et al., 2021). The recent launch of OpenAlex, an open index of scholarly outputs based on an aggregation of data from various sources such as Microsoft Academic, Crossref, ORCID, and Unpaywall (Priem et al., 2022), could potentially be leveraged for the assessment of various aspects of diamond OA, such as their distribution across language, disciplines, regions, income groups and authorships at a scale never studied before.

Despite the abundance of studies on gold and diamond journals, to our knowledge, none has investigated their respective coverage in commercial databases typically used for bibliometric research and research evaluation. This paper aims to address this gap by investigating the coverage of DOAJ journals (gold and diamond journals) in WoS and Scopus. Additionally, we seek to determine whether there is empirical support for the narrative that diamond journals are more community-driven and local in nature by analyzing the geographical and linguistic characteristics of gold and diamond journals and their relation to WoS and Scopus coverage. Further, it examines the characteristics of indexed diamond journals, such as country of publication, geographic concentration of authors, to produce a fuller portrait and subsequent understanding of how and why publications are represented in bibliometric databases. The specific research questions that we address are as follows:

1. What is the share of gold and diamond journals in the DOAJ?
2. What share of gold and diamond journals are indexed in WoS and Scopus?
3. How are diamond and gold journals distributed across regions, income groups, publishing languages, and disciplines?
4. What is the share of authorship in diamond and gold journals are indexed in WoS and Scopus?
5. How is authorship in diamond and gold journals distributed across regions, income groups and disciplines?

**Background**

**Open access models**

Open access is commonly understood as a funding or publishing model that aims to make research findings universally accessible at no cost to the reader. While the origins of the OA movement go back to the 1970s when scientists first started sharing papers on File Transfer Protocol (FTP) servers (Suber, n.d.), 2001's Budapest Open Access Initiative (BOAI) was the first significant operationalization of OA. The declaration proposed two main methods of OA dissemination: 1) self-archiving in an open electronic archive or repository, and 2) the creation of a new generation of OA journals along with a plan to help existing journals make the transition to OA (Budapest Open Access Initiative, 2002). Over the years, several OA models have been developed based on scientific publishers' OA practices (Eve, 2014; Piwowar et al., 2018; Suber, 2012), with Unpaywall's (Piwowar et al., 2018) classification being one of the most commonly used in recent years (Table 1). However, the Unpaywall classification does not make the distinction between diamond and gold OA.

Table 1. Unpaywall's Open Access models classification (Piwowar et al. 2018)

| Model | Definition |
| --- | --- |
| **Gold** | Published in an open-access journal that is indexed by the DOAJ |
| **Green** | Toll-access on the publisher page, but there is a free copy in an OA repository. |
| **Hybrid** | Free under an open license in a toll-access journal. |
| **Bronze** | Free to read on the publisher page, but without a clearly identifiable license. |
| **Closed** | All other articles, including those shared only on an ASN or in Sci-Hub |

*Gold open access and article processing charges*

Gold open access is now commonly accepted as an article published in an open access journal that requires an author fee (or article processing charge) for publication. These article processing charges were originally introduced by publishers to transfer the costs of publication from readers to authors, institutions, or funders to make all articles available in OA. Over time, publishers have been transitioning to publishing models based around APCs instead of subscription fees (Budzinski et al., 2020). Past studies have estimated gold OA APCs to range from $899 to $2,000 USD depending on the period of study and the data sources used (Crawford, 2021; Jahn & Tullney, 2016; Morrison, 2021; Siler & Frenken, 2020; Solomon & Björk, 2016). Butler et al. (2023) estimated that the five biggest publishers (Elsevier, Springer-Nature, Sage, Taylor & Francis, and Wiley) generated more than 1 billion USD in APC revenues between 2015 and 2018, including over 600 million in gold OA and 448 million in hybrid OA. Based on the DOAJ, Crawford (2021) estimated a total amount of APC revenue of around 1.27 billion USD for 2020, and a similar study by Zhang et al. (2022) provided estimates of about 2 billion USD for the year 2022. The French Ministry of Higher Education and Research has recently qualified the current market of APCs as unsustainable based on past costs and estimated future costs (Blanchard et al., 2022) and called for a transition to diamond OA (Matthews, 2021).

**Diamond open access**

Diamond OA is a publishing model in which papers are free for both authors and readers. They are often considered a subset of gold journals since they are technically compliant with the broad definition of gold (e.g., Piwowar et al., 2018). After several years of supporting gold OA, Science Europe, along with cOAlition S, OPERAS, and the French National Research Agency (ANR), have recently shifted their attention onto diamond OA with the aim of developing and expanding a sustainable, community-driven diamond OA scholarly communication ecosystem. The first step of this project consisted of a large-scale study in order "to gain a better understanding of the OA diamond landscape" (Bosman et al., 2021). Following the *Open Access Diamond Journal Study*, a 2022 workshop which gathered experts from the OA community led to the creation of a diamond OA Action Plan[1]. The main objective of the action plan is to "substantially increase the capacity of diamond journals to provide innovative, valid, reliable, and accessible publishing services", focusing on four main elements for its further development: (1) efficiency, (2) quality standards (3) capacity building, and (4) sustainability. More recently, the European Union-led Developing Institutional Open Access Publishing Models to Advance Scholarly Communication[2] (DIAMAS; 2022–2025) project was launched to "provide the research community with an aligned, high-quality, and sustainable scholarly communication ecosystem, capable of implementing Open Access as a standard publication practice across the European Research Area".

The *Open Access Diamond Journals Study* has shown that diamond journals represent a "wide archipelago of relatively small journals serving diverse communities" (Bosman et al., 2021, p. 7). According to their estimation, there are currently between 17,000 and 29,000 diamond journals in the scientific landscape, with about one third of them being indexed in the DOAJ. Diamond journals "struggle to be properly integrated into the ecosystem of scholarly publications" because they lack the leverage and the reputation to be included in major databases (Bosman et al., 2021). Additionally, diamond journals tend to publish fewer papers than APC-based ones, with their total number of papers going down over the years, coinciding with a rise in gold OA articles. Among all diamond journals published, North America and Asia have the lowest share, while Europe and South America account for more than two-thirds of all diamond journals. Social sciences and humanities have a high share of diamond journals compared to natural sciences and medicine. Additionally, the study reported that diamond journals are generally more multilingual than gold journals, with diamond journals serving mainly national authorship. Diamond journals also face a number of challenges in operations, including legal challenges, struggling to find reviewers, using unstable and potentially unsustainable platforms, monitoring and reporting due to their lack of indexation in more prestigious platforms and databases, and struggling with content visibility (Bosman et al., 2021). Regarding their sustainability, the authors highlighted that less than half of diamond journals reported breaking even, a quarter reported a loss, and others were not fully aware of their total costs. Costs of operations were generally low, with more than two-thirds declaring less than $10,000 USD in annual costs, and more than half of diamond journals relying on volunteer labor.

---

[1] https://www.scienceeurope.org/our-resources/action-plan-for-diamond-open-access/
[2] https://diamasproject.eu/

Using the complete DOAJ dataset from 2018, Siler and Frenken (2020) investigated the pricing of OA journals. According to their results, nearly three-quarters of journals listed in the DOAJ do not charge APCs, but diamond journals only published 43.2% of all OA articles published in the database, a situation also reported by the *Open Access Diamond Journals Study* (Bosman et al., 2021). In addition, only 10% of diamond journals had a Journal Impact Factor (JIF), highlighting the lower prestige of diamond journals. On the other hand, Butler et al. (2023) found that among 351,559 gold OA papers published between 2015 and 2018 by the largest academic publishers (Elsevier, Sage, Springer-Nature, Taylor & Francis and Wiley), only 12.4% were published in diamond journals, although it remains unclear if these articles are truly from diamond journals or the results of specific agreements with publishers. Their results highlight how diamond OA publishing is a phenomenon typically occurring outside the realm of large mainstream commercial publishers that dominate databases like WoS and Scopus.

**Open access disparities around the globe**

Fonturbel and Vizentin-Bugoni (2021) discuss how inequalities around the globe have been amplified through the shift to OA. While OA originally emerged as a response to scientific inequities, researchers have extensively discussed how low-income countries face barriers due to APC prices (Ellers et al., 2017; Klebel & Ross-Hellauer, 2022; Peterson et al., 2013; Santidrián Tomillo et al., 2022), leading to a lower uptake of gold journals in the Global South (Ezema & Onyancha, 2017; Mekonnen et al., 2021). Klebel and Ross-Hellauer (2022) show how the current incarnation of OA contradicts its goals through what they term the "APC-effect"; they correlated higher institutional resourcing with publications in journals with higher APCs, though variable across fields (Klebel & Ross-Hellauer, 2022). They found a distinct economic divide with high average APCs for countries with GDP per capita above $30,000 and heterogenous levels of average APCs for less wealthy countries, which they attribute to the effects of policies and alternative publishing models in Latin America and targeted research funding in Sub-Saharan Africa (Klebel & Ross-Hellauer, 2022). Furthermore, actions on the part of publishers, such as APCs waivers or discounts, have emerged as attempts to alleviate inequities caused by APCs (Lawson, 2015). Another unintended side effect of APC-based OA models includes the emergence of predatory or deceptive journals looking to capitalize on research through APCs (Beall, 2012).

Despite these barriers inhibiting uptake, the importance of OA for developing countries and the Global South has also been widely discussed in the scientific literature. For instance, by combining World Bank country income classification data, UNESCO's data on per capita gross national income (GNI) and bibliometrics data, Evans and Reimer (2009) discovered that paywalled articles were overwhelmingly cited by researchers in high-income countries, while OA papers tended to be cited more by developing countries, except where access to the internet and proper digital infrastructure remains a challenge. Similarly, Simard et al. (2022) showed that, on average, low-income countries are both publishing and citing OA literature at the highest rate, while publication practices varied considerably among high-income countries. They highlighted high OA adoption rates in Sub-Saharan African countries and low adoption rates in North Africa and the Middle East, a phenomenon also observed by Iyandemye and Thomas (2019) in the biomedical field. Additionally, they found a strong negative correlation between country per capita, income and the percentage of OA publication, with papers from international and inter-regional collaborations leading to higher OA rates than single-country papers (Iyandemye & Thomas, 2019). Looking into the factors associated with OA publishing, Momeni et al. (2023) showed that countries eligible for APCs waivers published more in OA, while those only eligible for discounts led to a lower ratio of OA publishing.

Another large-scale international journal study was completed using data from the open-source publishing platform Open Journal System (OJS; Khanna et al., 2022). Out of the 25,671 journals often excluded from major scientific indexes found in OJS, the vast majority were diamond OA, published in the Global South, and often operated in more than one language. Social sciences and STEM disciplines accounted for the largest proportion of journals, while humanities accounted for a lower share. The majority of OJS journals were found in middle-income countries, while less than 1% of journals were hosted in low-income countries. Only 1.2% of the OJS journals were found in WoS, while Scopus, Dimensions, OpenAlex, and Google Scholar indexed 5.7%, 54.3%, 63.8%, and 88.3% of them, respectively. Additionally, around one-fifth of the OJS journals were also indexed in the DOAJ. Investigating potential predatory publishers, they also found that 1% of journals using OJS were found in Cabell's Predatory Reports (Cabell, 2022) and a little over 1% in Beall's List (Beall, 2021). The authors argued that the narrative conflating predatory publishing with Global South journals and OA journals may be harmful to science because it "plays into the peripheralizing world system" without meaningfully addressing the problem (Khanna et al., 2022). Instead, they propose the creation of a set of publishing standards for the transparency of journal integrity, including publication facts such as publisher, discipline, rejection rate, number of reviewers, data availability, etc. (Khanna et al., 2022; Willinsky, 2022). Ultimately, they argue that scholarly publishing takes place on a far more global, diverse, and inclusive scale than popular databases suggest.

*On the coverage of bibliometric data sources*

The coverage of different bibliometric data sources has been subject to much scrutiny. Comparing the journal coverage of WoS and Scopus with Ulrich's extensive periodical directory, Mongeon & Paul-Hus (2016) showed that Scopus possesses a larger proportion of exclusive journals not indexed in WoS in all fields. Both WoS and Scopus were also found to overrepresent the natural sciences, engineering (NSE), and biomedical research (BM) and neglect the social sciences and humanities (SSH). English was also discovered to be overrepresented compared to other languages. Singh et al. (2021) compared the journal coverage of WoS, Dimensions, and Scopus and found Dimensions to be the most exhaustive in its indexing, while WoS and Scopus possessed similar coverage across different subject areas. Visser et al. (2021) provided a large-scale analysis of the coverage of scientific documents in Scopus, WoS, Dimensions, Crossref, and Microsoft

Academic and found Scopus's coverage was more exhaustive than WoS. They also observed that Microsoft Academic was the most comprehensive and covered more documents than the other databases (Visser et al., 2021). OpenAlex, a nascent open-source database based on Microsoft Academic Graph, has also recently inspired examination (Akbaritabar et al., 2023; Scheidsteger & Haunschild, 2023). Looking at various types of OA and comparing the percentage of indexed papers that are OA in the databases, Basson et al. (2022) found a higher percentage of papers indexed in Dimensions are OA compared to WoS, which proved to be particularly the case for papers originating from outside North America and Europe. They reasoned this is due to Dimensions' aim to index more broadly than WoS, including smaller national journals (Basson et al., 2022). These studies highlight different indexing choices (e.g., selective indexing vs. automated) made by platform holders and how these choices may ultimately affect research evaluation and bibliometric results based on their coverage of languages and disciplines.

## Data and methods

### DOAJ journals and their coverage in OpenAlex, Scopus, and Web of Science

We downloaded journal metadata from the February 2024 DOAJ Public Data Dump[3], the Scopus master journal list from the Scopus website, and the lists of journals indexed in WoS' Science Citation Index (SCI), Social Science Citation Index (SSCI), Arts and Humanities Citation Index (AHCI) and Emerging Sources Citation Index (ESCI), from the Clarivate Analytics Website. We decided to keep the ESCI separate from the WoS Core collection, as previous studies have traditionally excluded the ESCI from their analyses. We also retrieved all venues from a February 2024 snapshot of OpenAlex hosted by the Maritime Institute for Science, Technology, and Society (MISTS). We matched the DOAJ journals with those included in the other lists using the ISSN and the journal titles. The ISSN matching provided 20,021, 6,156, and 7,192 matches with journals from OpenAlex, WoS and Scopus, respectively. Since different journals occasionally share their names, we manually verified each journal based on their ISSN (physical or digital), publisher names, and cities of publication based on the information found in OpenAlex, the DOAJ, and on the ISSN website[4]. This resulted to an additional 735, 143, and 328 matches with OpenAlex, WoS and Scopus respectively.

### OA type

Since the DOAJ classification of gold journals and articles did not account for diamond OA, we created our own categories by separating OA journals and articles that charge APCs (gold) or not (diamond). While previous studies have estimated the total number of diamond journals to possibly as high as 29,000, it is important to consider that the majority of these journals have not been carefully examined for quality in the same way that the DOAJ journals have through the rigorous application process: some of those journals may require APCs that are not explicitly mentioned on their website, many of them do not self-identify as being a diamond OA or a no-APCs journal (Bosman et al., 2021).

---
[3] https://doaj.org/docs/public-data-dump/https://doaj.org/docs/public-data-dump/
[4] https://www.issn.org/

**Country, region, and income Group**

Classifying journals by country also has several potential implications related to their scope and target audience. The geographic location of a journal or a publisher may be exploited to gain credibility or to bury a less favourable branding. For instance, predatory publisher OMICS, who recently lost a $50 million USD ruling against the US Federal Trade Commission (FTC)[5] for the acceptance of thousands of articles with little to no peer review, was physically located in Hyderabad, India. After the introduction of a new brand, OMICS changed the URL and the aesthetics of their website, with various addresses scattered all over the world, including Spain, Belgium and the United Kingdom (Siler et al., 2021). Furthermore, in some cases, a legitimate publisher of a locally oriented journal may be geographically located in a different country or even continent, creating a situation where a journal publisher's location and the expected target audience do not match in our analyses. For example, in the DOAJ, the publisher for the South African Journal of Clinical Nutrition is listed as Taylor & Francis Group, United Kingdom, while the journal's website states, "Although the Journal is based in South Africa, it actively encourages articles from other African countries to act as a forum for the discussion of African nutritional issues"[6].

We mitigate the issue of the publisher country not necessarily reflecting the national roots or focus of journals by also considering in our analysis the countries (and their region and income group) of the institutions to which the authors publishing in a journal are affiliated, based on OpenAlex data. For each journal in OpenAlex, we collect all works published in or after 2015 and calculate the total number of authorships for each country so each journal can be represented by a distribution of author countries. We do this at the author-level (e.g., a paper authored by two Brazilian authors and one American would translate into two authorships for Brazil and one for the USA). The required metadata is not always available in OpenAlex, so we were able to obtain an authors' country distribution for 17,725 journals (about 86% of the journals included in OpenAlex).

We used the World Bank's country classifications to classify the journals based on their country's income level. Journals from Venezuela were not included in the income-level analysis since the country has been temporarily unclassified by the World Bank due to the unavailability of their data (World Bank, 2021). We further classified the journal publisher's countries by geographical regions using UNESCO's Regional Electoral Groups (UNESCO, 2022).

---

[5] https://www.ftc.gov/system/files/documents/cases/omics_ca9_ftc_answering_brief_10-11-19.pdf
[6] http://www.sajcn.co.za/index.php/SAJCN/about

**Disciplinary classification of journals**

For this paper, we assigned journals to one of three large research domains: Biomedical Research (BM), Natural Sciences and Engineering (NSE), and Social Sciences and Humanities (SSH). This was done using several data sources to gather information about the topical or disciplinary focus of a journal. One of these data sources was the Science-Metrix journal classification (Archambault et al., 2011), which assigns journals indexed in Scopus to 6 domains further divided into fields and subfields. We also associated a Science-Metrix domain with the journals based on the most cited domain from their reference lists. Additionally, we used the journals' keywords and subjects included in the DOAJ metadata. The final assignment of a discipline to a journal was done manually. Overall, we assigned a domain to all but two of the 19,238 journals (5,058 in BM, 5,448 in NSE, 10,640 in SSH). The sum of these numbers is greater than the total number of journals because 1,908 journals were assigned to more than one domain. Two multidisciplinary journals were not classified.

**Final dataset**

All the data sources mentioned above were combined to constitute the dataset used in this study, which comprises the variables listed in Table 2.

Table 2. List of variables used included in the final dataset.

| Variable | Description | Source |
|---|---|---|
| publisher_country | The country where the publisher of the journal is located. | DOAJ |
| publisher_region | The region of the publisher country. | World Bank |
| publisher_income_group | The income group of the publisher country. | World Bank |
| authors_country | Distribution of authors' affiliation country for the journal. | OpenAlex |
| authors_region | Distribution of authors' affiliation region for the journal. | OpenAlex |
| authors_income_group | Distribution of authors' affiliation income group for the journal. | OpenAlex |
| language | List of all languages in which the journal accepts and publishes works. | DOAJ |
| domain | One of three large research areas: Biomedical Research (BM); Natural Sciences and Engineering (NSE); Social Sciences and Humanities (SSH). | Science-Metrix Classification + DOAJ + OpenAlex + Manual classification |
| oa_type | Two types based on the APC amount: Gold (APC > 0) and Diamond (APC = 0). | DOAJ |
| is_in_Scopus | The journal was found in the Scopus journal list | DOAJ + Scopus Master Journal list |
| is_in_WoS | The journal was found in the Web of Science core collections (SCI-E, SSCI, A&HCI) | Web of Science Master Journal list |
| is_in WoS-ESCI | The journal was found in the Web of Science Emerging Sources Citation Index (ESCI) | Web of Science Master Journal list |
| is_in_OpenAlex | The journal was found in OpenAlex | May 2022 OpenAlex data dump. |

## Results

**Coverage of diamond open access journals in OpenAlex, WoS and Scopus**

Looking at the DOAJ journals overlap between the three databases (Figure 1), we see that for both diamond and gold, OpenAlex indexes almost the entirety of WoS and Scopus journals. However, diamond is where OpenAlex truly shines, with over 12,500 journals indexed, including 60% of all diamond OA journals that are not found in either WoS or Scopus. In comparison, Scopus and WoS respectively index only 99 and 30 diamond journals that cannot be found in other databases. Over half of the diamond journals indexed in Scopus can also be found in WoS, while more than two thirds of the diamond journals in WoS can also be found in Scopus. OpenAlex indexes over 6,500 gold journals, with about one third of them not being found in the legacy data sources, compared to only 21 and for Scopus and WoS. Nearly half (45%) of all gold journals in the DOAJ can be found in all three databases.

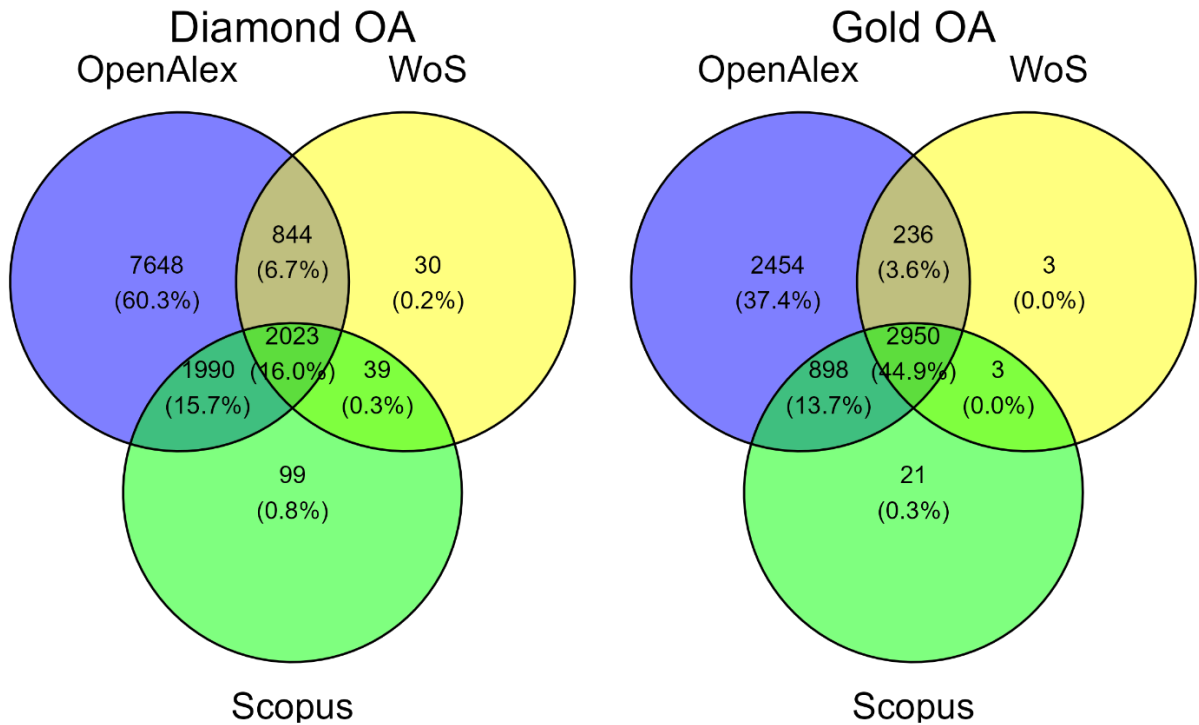

Figure 1. Overlap of DOAJ diamond and gold journals in OpenAlex, Scopus and Web of Science (including the Emerging Sources Citation Index).

Figure 2 shows the coverage of DOAJ diamond and gold journals in OpenAlex, and their coverage by WoS and Scopus. Our results show a generally good coverage of gold BM (77%) and NSE (68%) journals by the traditional databases. However, less than half of BM (47%) and NSE (45%) diamond journals are indexed in the traditional databases. Additionally, our results indicate an underwhelming coverage of the SSH, with the majority of these journals (70% gold, 68% diamond) not being indexed in either WoS or Scopus.

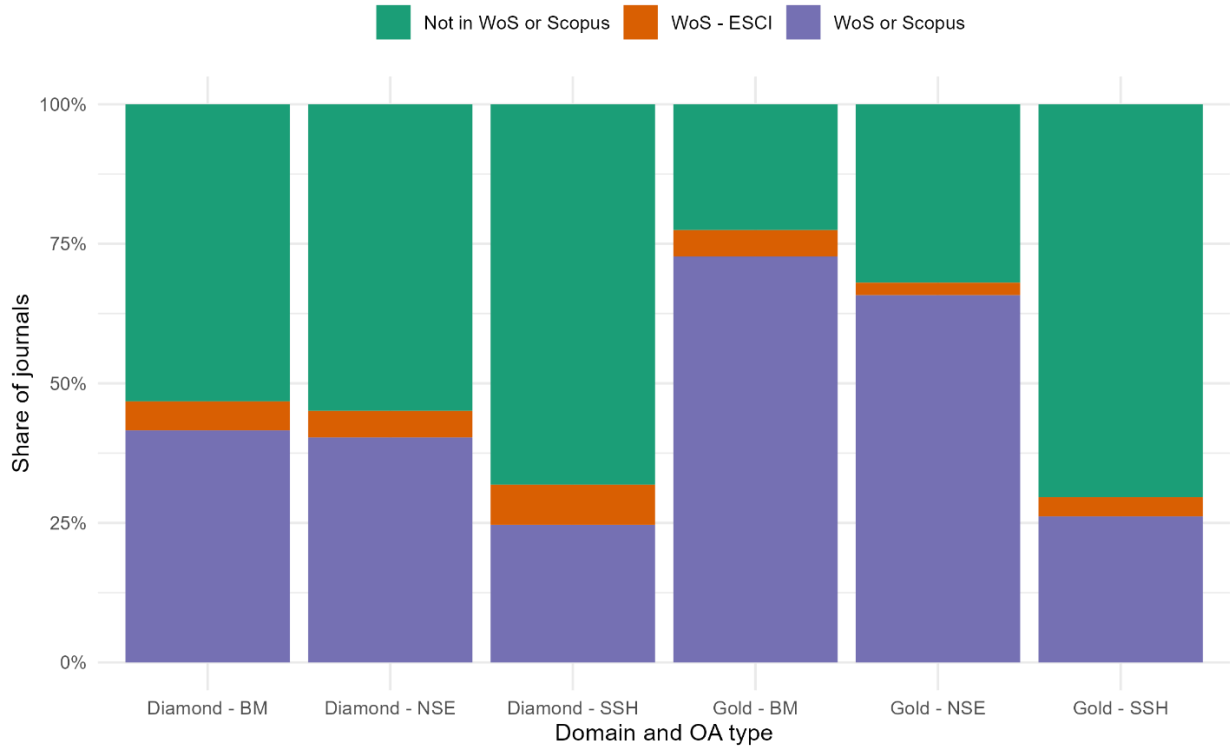

Figure 2. Coverage of DOAJ journals in Web of Science and Scopus by discipline

**Geographic coverage from the publishers' perspective**

*Regions*

Figure 3 illustrates the share OA of journals by publisher region, and domain. We find that most gold journals are in Europe & Central Asia, East Asia & Pacific, as well as North America for every domain. Additionally, nearly two thirds of gold journals are indexed in the proprietary databases, as opposed to about slightly over a third for diamond journals. Our results also highlight the importance of diamond journals for the Latin America & Caribbean region: while this region hosts less than 2.5% of gold journals, it accounts for nearly a quarter of diamond journals, ranking second only behind Europe & Central Asia, with most of their journals not being available in WoS or Scopus. We also find that SSH OA journals tend to be underrepresented in the databases, while the landscape of NSE and BM journals is generally similar with gold journals generally being indexed, while diamond are mostly underrepresented in traditional databases outside of Europe & Central Asia.

Looking at the countries with the highest share of journals, we find that, apart from gold NSE and BM journals, these journals are generally not very well covered by the traditional databases, especially in Indonesia and Brazil. Great Britain, the United States, and China currently dominate the gold OA landscape, hosting nearly two-thirds of BM journals, with nearly all of them being indexed in traditional databases, while diamond BM journals are mostly split among non-Western countries such as Iran, Brazil, and India. Our results also show a vast difference between the gold and diamond journal landscape. For instance, a small group of mostly richer countries (Indonesia, Great-Britain, China and The United States) control more than half of the gold journal landscape, while the diamond landscape is much more evenly divided among a variety of diverse countries. The landscape of SSH journals, especially for gold OA, is dominated by Indonesia which hosts more than a third of all journals. However, nearly all these journals are excluded from the traditional databases. Brazil hosts the highest number of diamond SSH journals, with most of them being outside of traditional databases.

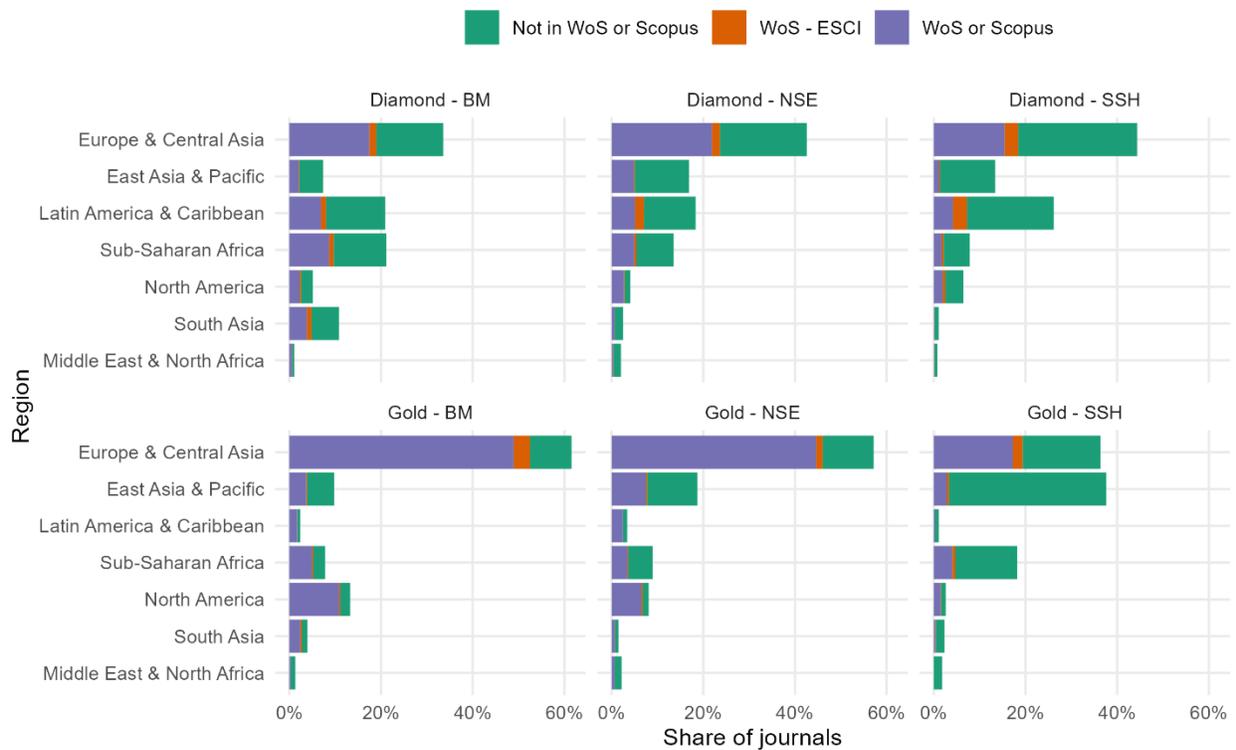

Figure 3. Share of journals by publisher region, domain, and OA type.

To further examine the differences in the publisher country representation based on domain, journal type, and coverage, we calculated the Gini coefficient for each group. The results are presented in Table 3. The data shows strong concentration patterns in general but slightly lower Gini coefficients for diamond journals compared to gold journals across domains.

Table 3. Gini coefficient for publisher countries by domain, type of journal, and coverage.

| Domain | OA Type | OpenAlex only | WoS - ESCI | WoS or Scopus |
|---|---|---|---|---|
| BM | Diamond | 0.908 | 0.938 | 0.918 |
|  | Gold | 0.946 | 0.971 | 0.962 |
| NSE | Diamond | 0.904 | 0.917 | 0.895 |
|  | Gold | 0.952 | 0.961 | 0.950 |
| SSH | Diamond | 0.909 | 0.938 | 0.906 |
|  | Gold | 0.960 | 0.958 | 0.944 |

*Income groups*

Aggregating the share of journals by income group (Figure 4), we see that high income countries have the highest share of gold OA journals, with vast majority of those journals being indexed in traditional databases. Over half of gold SSH journals come from lower middle income countries, but most of them are not available in WoS or Scopus. However, looking into diamond OA journals tells a completely different story. Publishers from upper middle income countries have the highest number of diamond journals, with a share 40% or more in every domain. Unsurprisingly, these journals are underrepresented in traditional databases, especially in SSH. Additionally, every income group outside of high income has less than half of their diamond journals indexed in WoS or Scopus.

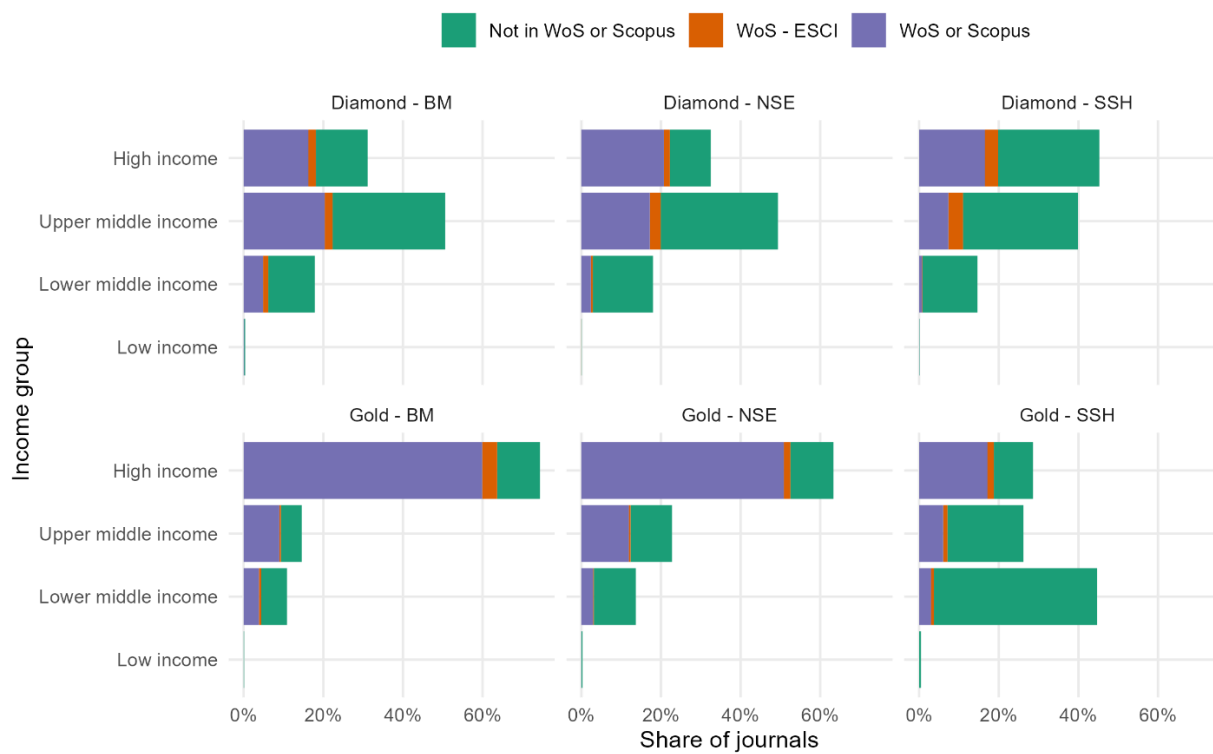

Figure 4. Share of journals by publisher, country income group, domain, and OA type.

**Linguistic coverage**

Figure 5 shows the breakdown between English and non-English journals. The multiple language category consists of journals that accept submissions in both English and at least one other language. Overall, we see that English-only journals are the norm for BM and NSE gold journals accounting for more than three-quarter of all journals indexed, but that the landscape of languages is very diverse in diamond journals, as well as gold SSH journals. Unsurprisingly, English-only journals are overwhelmingly more indexed in WoS and Scopus compared to non-English or multiple languages journals, especially for gold journals.

The share of strictly non-English gold journals varies between 4% for BM and 20% for SSH, with most of them not being available in the tradition databases. Looking at diamond journals, the language breakdown is more evenly divided among journals in both overall number of journals, but also among the three different domains. Diamond OA SSH journals tend to be more multilingual (49%) or non-English (27%) than both NSE and BM journals, where nearly half of the journals are in English only. Both multilingual and non-English diamond journals also tend to be less indexed than diamond OA English journals in every discipline. Looking at specific languages, Indonesian is generally the most common second language accepted for submission for gold journals, while Spanish and Portuguese are among the most frequent for diamond journals.

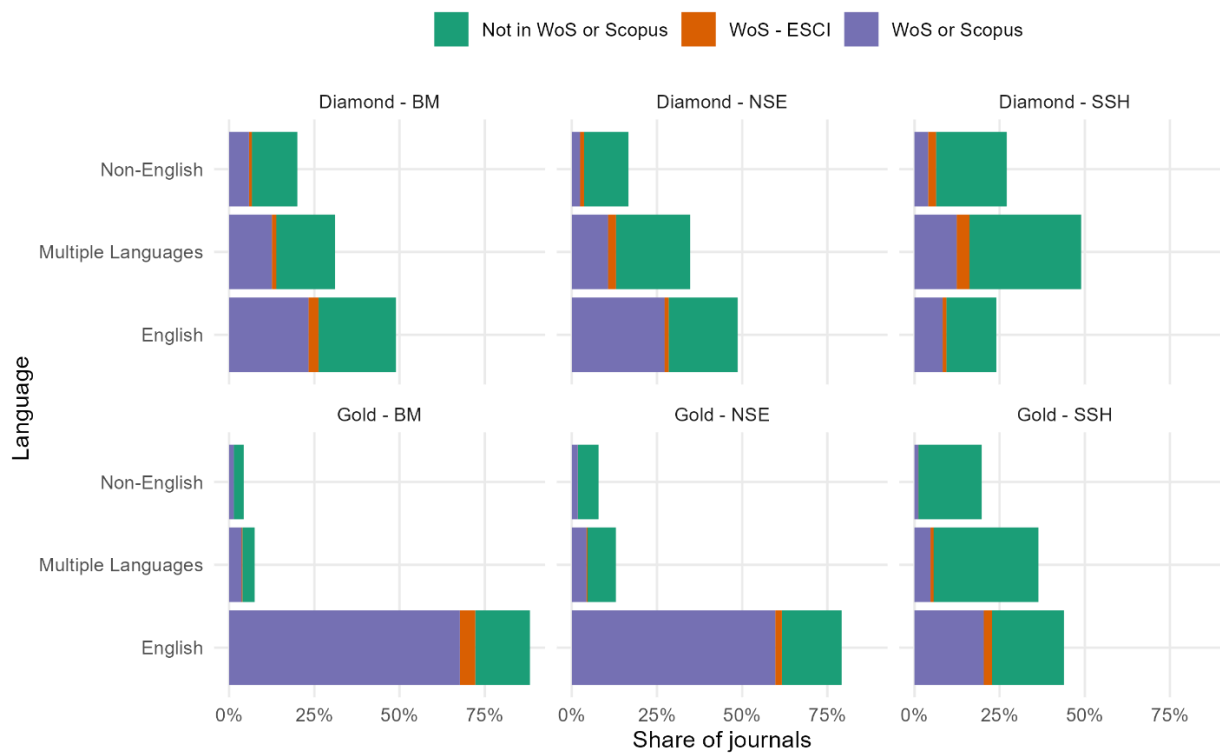

Figure 5. Share of journals by language, domain, and OA type.

**Geographic coverage from the authors' affiliations perspective**

*Regions*

When aggregating numbers at the region level (Figure 6), we observe the strength of Latin America & the Caribbean authorships in diamond OA journals, but with lower coverage in WoS or Scopus compared to other regions, especially in SSH. Europe & Central Asia has the overall largest share of authorship, while East Asia & Pacific has the highest share of authorship in gold journals, which coincide with the strong culture of gold OA in Indonesia that we will discuss later. Among gold journals in SSH, nearly one third of authors were linked to articles not indexed in WoS or Scopus. Diamond SSH authorship, especially in Latin America & the Caribbean and East Asia & Pacific, was mostly affiliated with articles not appearing in the two proprietary databases. In NSE and BM, nearly all authors of gold OA articles were linked to WoS or Scopus, while less than a quarter of their authors in gold OA journals were not indexed in the proprietary databases.

At the country-level, Indonesia appears to have the largest number of authorships in gold OA in SSH, but the vast majority in journals not indexed in WoS or Scopus. In terms of authorships in diamond OA journals, Brazil has the largest share of authorships in BM, coming second behind China in NSE, and being by far the most the dominant country for diamond OA in SSH, where several other Latin American & Caribbean countries also appear among the top countries.

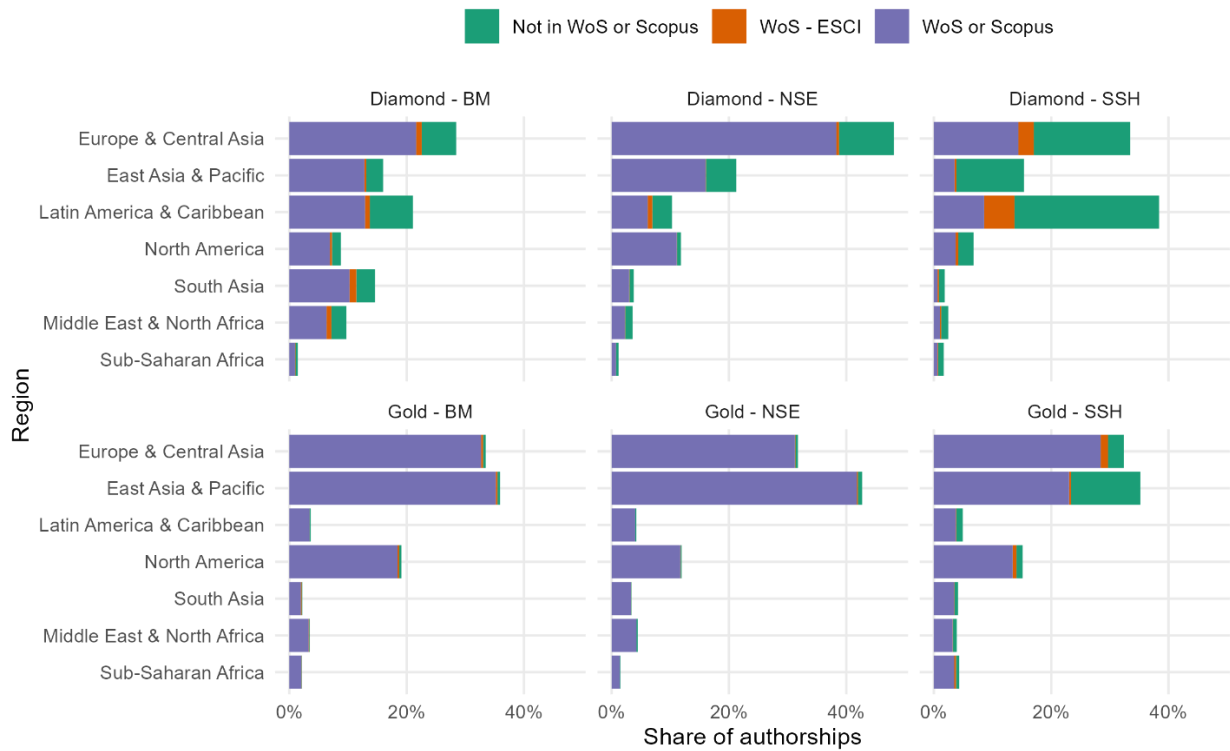

Figure 6. Distribution of authorships in gold and diamond OA journals indexed in the DOAJ across region and their coverage in WoS and Scopus (2015-2022).

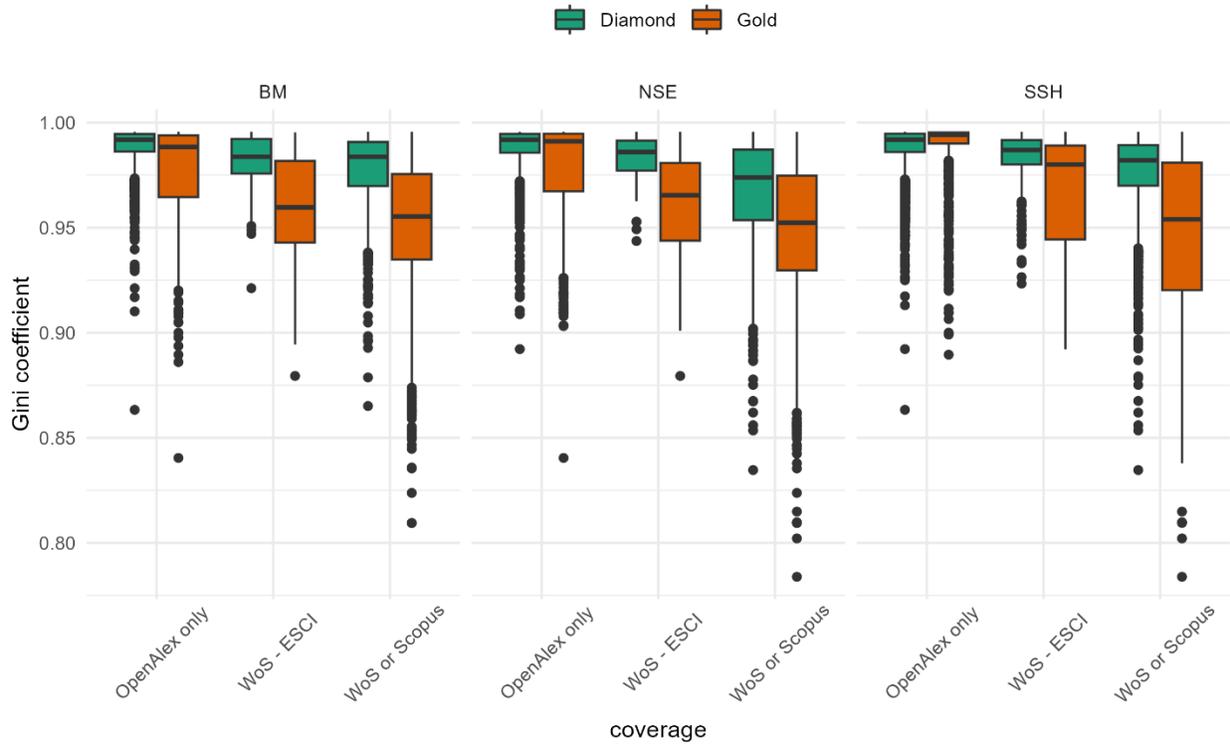

Figure 7. Distribution of Gini coefficient of authorships across countries at the journal level by type and coverage (A = OpenAlex only, B = WoS - ESCI, C = WoS or Scopus) in each domain.

Figure 7 displays the distributions of the Gini coefficients of the geographic distribution of authorships across countries at the journal level by OA type, database, and domain. This allows us to determine whether gold OA journals tend to have a geographically broader authorship than diamond OA journals, or vice-versa. Because the distributions are skewed, we performed a Kruskal-Wallis test for each domain, a non-parametric alternative to the one-way ANOVA, to establish that there exist statistically significant differences between the groups formed by crossing the journal type and the coverage. We also performed pairwise comparisons using the Wilcoxon rank sum test with Benjamini-Hochberg adjusted p-values. The Kruskall-Wallis tests showed statistically significant differences between the groups in BM ($\chi^2$ = 34,307.63, df = 5, p < 0.001), NSE ($\chi^2$ = 26728.51, df = 5, p < 0.001), and SSH ($\chi^2$ = 34891.11, df = 5, p < 0.001). All pairwise Wilcoxon comparisons were statistically significant.

*Income groups*

Aggregating authorship at the country income group level (Figure 8) show that high income countries have the highest share of authorship in every domain and type of journal apart from diamond journals in SSH where upper middle income countries overtake their counterpart.

In SSH, authors affiliated with high income countries account for nearly half of all authors, with most of them being affiliated with WoS or Scopus. SSH authors from lower income countries are strongly underrepresented in traditional databases, especially in diamond journals, with more than half of their authorship being outside of the traditional databases. Most authors of diamond OA SSH papers from all regions were not associated with WoS or Scopus in every income group, except for authors from high income countries. In NSE, high income countries account for over half of authorship, while authors from upper middle income, lower middle income and low income countries respectively account for 38%, 9%, and less than 1% of authorship. Nearly all NSE authorship in gold OA articles is associated with traditional databases, compared to 79% for diamond OA articles. In gold journals, high and upper middle countries dominate authorship in the biomedical domain with over 90% of all authorship, including nearly all of them being indexed in traditional databases. However, authorship in diamond OA BM articles is much more evenly distributed among middle income and high-income groups. Low income countries have a minimal share of authorship across all domains and journal types.

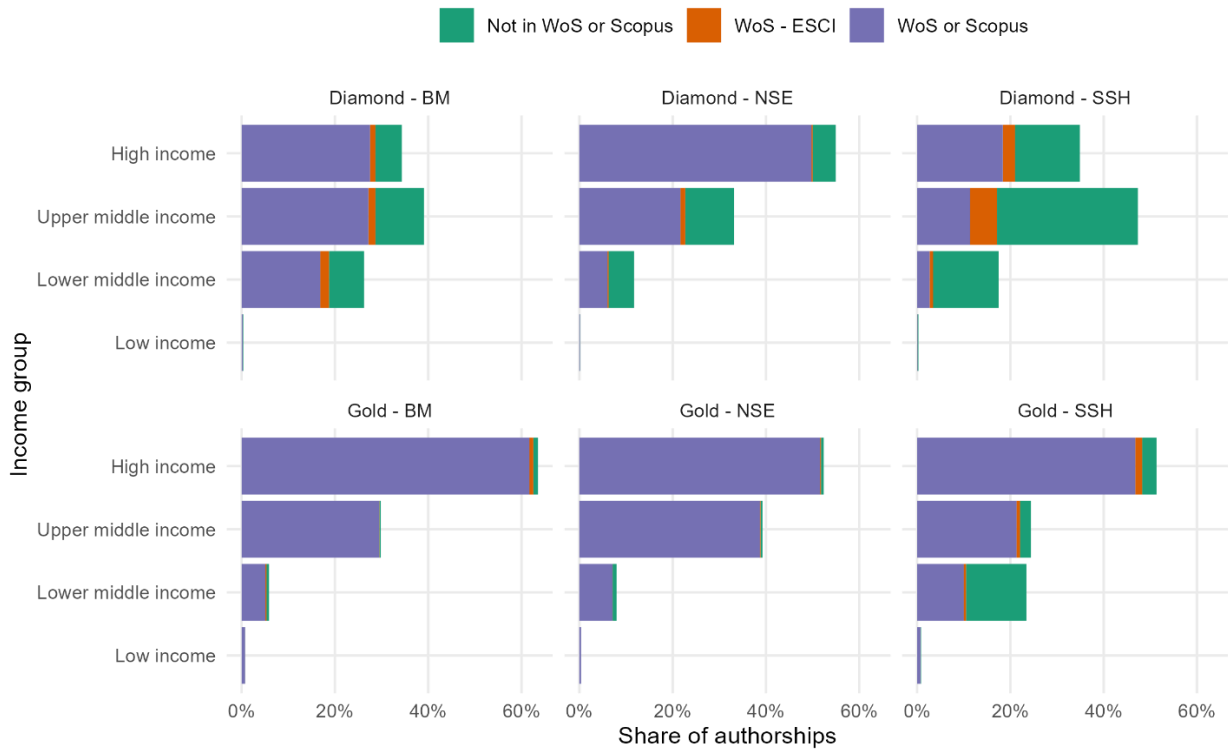

Figure 8. Distribution of authorships in gold and diamond OA journals indexed in the DOAJ across income group and their coverage in WoS and Scopus (2015-2022).

## Discussion and conclusion

Previous studies have estimated the total number of diamond journals to possibly be as high as 29,000 (Bosman et al., 2021). However, most of these journals have not been carefully examined for quality in the same way that the DOAJ journals have through their rigorous approval process, making the DOAJ-OpenAlex combination used in this analysis a unique benchmark. Unsurprisingly, most journals included in the DOAJ are also included in OpenAlex, while WoS and Scopus index less than half of them. WoS's coverage of diamond journals is particularly low, especially when you remove the journals covered in the WoS core collection, which is concerning given that most studies using WoS as a data source generally only use the Core indexes.

On the other hand, while the DOAJ indexes more diamond than gold journals, the number of authors associated with gold articles considerably outweighs those associated with diamond journals. Diamond journals also tend to publish less articles, accept manuscripts in languages other than English, to be smaller in scope, and to serve diverse geographic or disciplinary communities (Bosman et al., 2021; Khanna et al., 2022; Siler & Franken, 2020), which reinforces the need to use more inclusive databases when investigating OA in order not to perpetuate the exclusion of already underrepresented research communities (Basson et al., 2022). This seems to corroborate previous claims that diamond journals are more local and community-oriented and may not have the same publishing capacities as gold journals (Bosman et al., 2021; Khanna et al., 2022; Siler & Frenken, 2020).

Overall, our results show significant differences in the distribution across countries, geographic regions, and country income levels between gold and diamond journals, with diamond journals displaying more even, although still quite heavily skewed, distributions. We also find that DOAJ journals in the NSE and BM are much better covered by WoS and Scopus than SSH journals, which aligns with past findings from Mongeon and Paul-Hus (2016). Interestingly, there appear to be important coverage differences between gold and diamond journals in NSE and BM but not in SSH. Indonesia and, Latin America & the Caribbean countries are among the countries with the most OA journals. The higher number of OA journals (especially diamond) in Latin America & the Caribbean may be explained in part by regional initiatives and platforms such as SciELO, Amelica, and Redalyc. As for Indonesia, their entire scholarly communication ecosystem is directly embedded in universities, meaning that they mostly operate from university funding (Matthias, 2018), but they must also be exclusively available in OA after Indonesia passed a 2019 law on National Knowledge System and Technology that mandates OA licenses for research outputs to guarantee public access to research[7].

Unsurprisingly, English was the most accepted language for submission in diamond and gold journals in all domains. While Indonesian was the second most commonly accepted language for submission in gold journals, Spanish and Portuguese, respectively, came second and third in every domain for diamond journals, once again highlighting a potential "SciELO effect." Our results concerning journal publication language have also shown the importance of diamond OA for local communities. While gold journals, especially those indexed in WoS or Scopus, mostly accept article submissions in English, nearly half of diamond DOAJ journals in all fields consider accepting submissions in languages other than English. Furthermore, a significant share of diamond journals do not publish research in English at all. Interestingly, a good share of journals (30%) that publish in languages other than English are indexed in WoS or Scopus, although journals that publish in English have a much better coverage. Our findings thus illustrate how diamond journals may serve as publishing venues for a more geographically and linguistically diverse population of researchers and, plausibly, for more local research, in line with previous findings.

Just like the several articles that have previously studied the coverage of bibliographic databases (Akbaritabar et al., 2023; Basson et al., 2022; Mongeon & Paul-Hus, 2016; Priem, 2022; Scheidsteger & Haunschild, 2022; Singh et al, 2021; Visser et al., 2021), our results also have implications for research evaluation and monitoring compliance to OA policies, particularly journal-based OA policies. First, there is a clear Global North bias in terms of coverage of OA journals in WoS and Scopus. As demonstrated by Basson and their colleagues (2022), the choice of a specific data source may have a direct influence in calculating OA rates and compliance with OA policies, especially for researchers in SSH or from the Global South, where there are smaller and more local and national journals. However, the use of more inclusive data sources such as OpenAlex and Dimensions for research evaluation may also lead to other types of challenges. While traditional data sources rely on strict indexation methods and publishers' data, OpenAlex relies on multiple data sources (i.e., Crossref, Microsoft Academic Graphs, ORCiD, etc.), which may sometimes lead to incomplete or problematic metadata, especially for author disambiguation and affiliations (Priem et al., 2022), which once again may lead to better coverage of articles from the Global North because of more complete and uniform metadata structure.

**Limitations**

As previously discussed, classifying journals by country poses several potential limitations and carries implications related to their scope and target audience. For instance, journals or publishers in particular geographic locations may be wielded as a means of gaining credibility. This geographic classification may also present conflict with analyzing publishing locations and target audiences since they may not always accurately represent their true scope (e.g., A South African journal hosted by a European publisher). Other limitations include the exclusion of journals not included in the DOAJ and potential incomplete metadata, mainly in terms of authors' affiliations. Our classification of diamond journals may also lead to certain issues. While we define diamond OA as journals that are free for both readers and authors based on the APC information in the DOAJ, this definition does not necessarily take into consideration the community and non-profit aspects of diamond OA. As previous demonstrated, several journals published by major for-profit publishers could technically be defined as diamond OA since they do not charge readers or authors (Butler et al., 2023; Simard et al., 2024). However, these journals remain for-profit journals where the production costs (and profits) may be covered by a scientific society that the journal serves. Journals may also be "temporarily" in specific contexts, such as a promotional attempt to attract readers and authors by temporarily removing reading or author fees (or both) or simply by removing paywalls and APCs for COVID-related journals and articles.

---

[7] https://peraturan.bpk.go.id/Home/Details/117023/uu-no-11-tahun-2019.

**Concluding remarks**

Ultimately, our study suggests that APC-based OA models are not the only option for researchers to publish and disseminate their work in OA journals; in fact, most OA journals indexed in the DOAJ do not rely on APCs to operate. Tools such as the DOAJ and OpenAlex provide a certain gateway to bypass the lack of coverage of diamond journals in more selective databases such as WoS and offer greater visibility to such journals. These databases can also act as a mechanism for more inclusive research evaluation that accounts for diamond OA and local journals instead of being limited to "top journals" indexed by WoS or Scopus. The latter two databases are often viewed as quality gauges for scholarly journals in research evaluation (e.g., grants, tenure track, etc.), especially through their indicators (e.g., Journal Impact Factor), obstructing a fuller picture of a scientific research system. Supporting the growth of such journals could be facilitated by directing the costs used to subscribe to for-profit journals and to pay for exorbitant APC prices, reinjecting them into the operating costs of diamond journals (Fontúrbel & Vizentin-Bugoni, 2021) models, similar to what the DIAMAS project aims to do. This would also require continued social change among the research community through recognizing and legitimizing the value of diamond journals. More inclusive research assessment practices are key, but directly choosing to submit work to them, cite work published in them, as well as contribute to their workload (through peer-review and editorial board membership) would potentially lead to their increased use and rise in credibility. To achieve this goal, the full extent of the research community must be leveraged. The scholarly communication system is an interconnected and interdependent network in which all actors (e.g., researchers, librarians, publishers, and funding and policy bodies) have the capacity to influence and enact a transition from a subscription and APC-based publication to diamond OA and a fully open access global research ecosystem (Greussing et al., 2020).

**Competing interests:** Vincent Larivière is Editor-in-Chief of Quantitative Science Studies.

**Acknowledgments:** Marc-André Simard acknowledges funding from FRQSC - Doctoral training scholarship and Centre interuniversitaire de recherche sur la science et la technologie (CIRST) International Mobility Program.

Authors Contributions
Conceptualization: MAS, PM, VL
Data curation: MAS, MH, PM
Funding acquisition: MAS, VL
Investigation: MAS, PM
Methodology: MAS, PM
Project administration: MAS, VL, PM
Supervision: MAS, VL, PM
Visualization: MAS, PM
Writing – original draft: MAS, IB, MH, PM
Writing – review & editing: MAS, IB, MH, PM, VL